\documentclass{PoS}

\title{27-day Variation of the Three Dimensional Solar Anisotropy of Galactic Cosmic Ray: 1965-2014
}

\ShortTitle{27-day Variation of the 3D Anisotropy of Cosmic Rays}

\author{\speaker{R. Modzelewska }\thanks{A footnote may follow.}\\
        Inst.  Math. and Physics, Siedlce University, Siedlce, Poland\\
        E-mail: \email{renatam@uph.edu.pl}}
\author{M. V. Alania  \\
        Inst.  Math. and Physics, Siedlce University, Siedlce, Poland\\
        Institute of Geophysics, Tbilisi State University, Tbilisi, Georgia\\
       E-mail: \email{alania@uph.edu.pl}}

\abstract{The temporal evaluation of the 27-day variation of the three dimensional (3D) galactic cosmic ray (GCR) anisotropy has been studied for 1965-2014. 3D anisotropy vector was obtained based on the neutron monitors and Nagoya muon telescopes data. We analyze the 27-day variation of the (1) two dimensional (2D)  GCR anisotropy in the ecliptic plane, and (2) north-south anisotropy  normal to the ecliptic plane. Studying  the time line of the 27-day variation of the 2D GCR anisotropy, we confirm that the average amplitude in the minimum epoch of solar activity is polarity dependent, as it is expected from the drift theory. The amplitude in the negative polarity epochs is less  as we had shown  before. The feeble 11-year variation connected with solar cycle and strong 22-year pattern connected with solar magnetic cycle is visible in the 27-day variation of the 2D anisotropy for 1965-2014. We show that the 27-day variation of the $GG$ index (being a measure of the north-south asymmetry) varies in accordance to solar cycle with a period of 11-years, being in good agreement with the 27-day variation of the $At$  component of the GCR anisotropy calculated by the IZMIRAN group. Detailed analysis are presented for the solar minimum 2007-2008 of solar cycle no.23  and solar maximum 2013-2015 of solar cycle no 24. In the solar cycle no. 24 $GG$ index, calculated by Nagoya telescopes data, is highly anticorrelated with $By$ component of the interplanetary magnetic field (IMF) and shows a clear recurrent changes related to the Sun's rotation.}

\FullConference{The 34th International Cosmic Ray Conference,\\
		30 July- 6 August, 2015\\
		The Hague, The Netherlands}

\begin{document}

\section{Introduction}
An average solar diurnal anisotropy of galactic cosmic rays (GCRs) is formed due to modulation of GCR particles by the solar wind. It has been explained based on the diffusion-convection theory of GCR propagation in the heliosphere \cite{Ahluwalia62}-\cite{Parker64}, as a consequence of the equilibrium established between the radial convection of GCR particles by solar wind and the inward diffusion of GCR particles (owing to an existence of radial gradient)  along the interplanetary magnetic field (IMF). Despite, three-dimensional (3D) anisotropy vector determines the distribution of the stream of cosmic rays in the 3D heliosphere, nevertheless,  the complete nature of the 3D  GCR anisotropy from top to bottom is not  understood  yet. Scientific staff of IZMIRAN's  cosmic ray laboratory\\
(http://helios.izmiran.troitsk.ru/cosray/main.htm), have calculated the components  $Ar$, $Af$, and $At$ of the 3D anisotropy by Global Spectrographic Method (GSM) \cite{Krymsky66}, \cite{Krymsky67} based on hourly data from all operating neutron monitors (NMs).  Unfortunately, the derivation of the $At$   component is possible  with an accuracy up  to  constant, and so,  a value of the north-south anisotropy $At$   obtained by GSM method is not accurate $[http://cr20.izmiran.rssi.ru/AnisotropyCR/index.php]$. That is, sorrowfully, a deficiency  of the GSM related to the nature of NMs data. In general determination of the two dimensional (2D) anisotropy in the ecliptic plane is feasible based on establishing the radial $Ar$ and tangential $Af$ components by the harmonic analysis method for an individual detector (e.g., NM or Muon Telescope (MT)). Alania et al., \cite{Alania08} show that GSM gives the same results as the individual NM with cut-off rigidity $< 5$ GV.

Swinson \cite{Swinson69} first suggested that the north-south anisotropy could be related to the cosmic ray flow caused by a positive heliocentric radial density gradient $Gr$ of cosmic rays and the $By$ component of the interplanetary magnetic field (IMF) $B$, as expressed by the vector product, $By$ x $Gr$. Since the average  $By$ and $Gr$ lie in the ecliptic plane, the direction of the vector product $By$ x $Gr$  (or the north-south anisotropy) is expected to be perpendicular to the ecliptic plane and to reverse direction with changes of the magnetic field direction. It is directed upward  when $By$ is positive (IMF away from the Sun-$A>0$) and vice versa when $By$ is negative (IMF toward the Sun-$A<0$).
Papers \cite{Bieber86}- \cite{Bieber93} studied the cosmic ray anisotropy vector in three dimensions using data from polar located NMs.

An alternative method to study north-south anisotropy is proposed by Mori and Nagashima \cite{MN79}. They have introduced an index $GG$ calculated from Nagoya MT data as: $GG = \frac{1}{2}[(49N-49S) + (49N-49E)]$. The terms represent the counting rate differences for telescopes pointing in North, South, and East directions at $49^{0}$ zenith angle. The $GG$ index is free of noise in isotropic intensity caused by Forbush decreases, periodic variations, atmospheric temperature effects, and geomagnetic cut-offs. $GG$ index mainly reflects change of difference between the intensity from north polar direction and that from north and  parallel to the equatorial plane directions. Counting rate differences of the north-south and north-equatorial could  not be precisely contained  the same type of information. In spite, a $GG$  is accepted by cosmic ray community as a good alternative index to study the north-south asymmetry of the cosmic ray flux e.g., \cite{Munakata14}.

The large amount of papers dedicated to the recurrent variations of cosmic rays related to the period of the solar rotation, further in this paper called 27-day variation,  are devoted to the 27-day variation of the GCR intensity. The 27-day variation of the GCR anisotropy in general has been studied less intensively up to present. It partially is connected with the small amplitudes of the  GCR anisotropy ($< 0.3\%$, measured by NMs) and with a large dispersion comparable with the accuracy of hourly data of NMs. The existence of the 27-day recurrence of the GCR anisotropy was first reported by Yoshida and Kondo \cite{YK39} and later confirmed by others.
The 27-day variation of the north-south anisotropy was studied in the series of papers by Swinson and coauthors \cite{SY92}-\cite{SF95}. They showed that 27-day variation of the north-south anisotropy is correlated with solar activity and this correlation is not clearly dependent upon solar magnetic polarity. Recently, Yeeram et al., \cite{Y} studied recurrent trains of enhanced GCR anisotropy under influence of corotating solar wind structures near the recent solar minimum. They have noted  that during the recurrent time ($\sim27$ days) the GCR anisotropy  was often enhanced or suppressed in accord with the sign of the interplanetary magnetic field $B$, possibly demonstrating   a contribution from a mechanism involving a southward gradient $Gt$ in the GCR density, and an anisotropy caused  by drift, $B$x$Gt$. However, to be sure in this  suggestion there is necessary to carry out  an additional quantitatively  analyze involving  data of NMs with low cut-off rigidities ($<5$ GV).

Our aim in this paper is to study long-term changes of the 27-day variation of the GCR anisotropy for 1965-2014. Detailed analysis are presented for the solar minimum 2007-2008 of solar cycle no.23 and for the solar maximum 2013-2015 of solar cycle no. 24.

\begin{figure}[tbp]
  \begin{center}
\includegraphics[width=0.8\hsize]{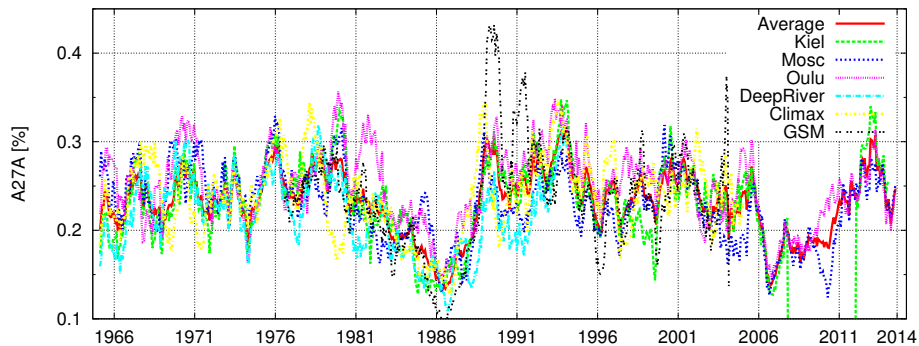}
\end{center}
\caption{\label{fig:1} Time line of the amplitudes of the 27-day variation of the 2D GCR anisotropy (A27A), smoothed over 13 Sun's rotations, for Climax, DeepRiver, Kiel, Moscow and Oulu individual NMs and for all combining NMs by GSM method for 1965-2014.}
\end{figure}

\begin{figure}[tbp]
  \begin{center}
\includegraphics[width=0.8\hsize]{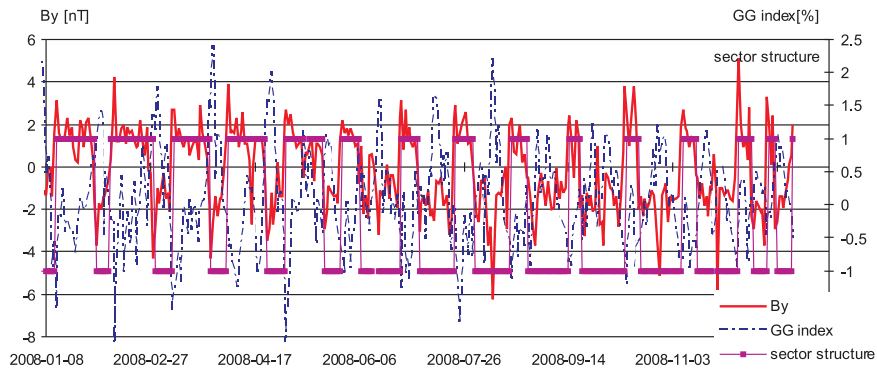}
\end{center}
\caption{\label{fig:2} Daily $GG$ index, $By$ component and the sector structure of the IMF in 2008. }
\end{figure}

\begin{figure}[tbp]
  \begin{center}
\includegraphics[width=0.85\hsize]{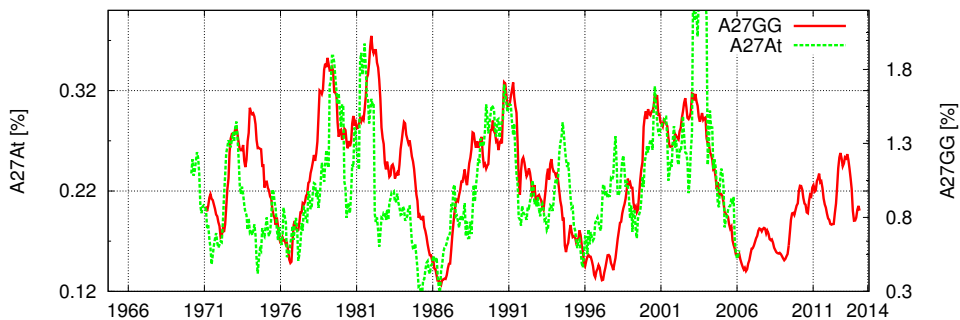}
\end{center}
\caption{\label{fig:3} Temporal changes  of the smoothed over 13 Sun's rotations amplitudes of the 27-day variation of the $GG$ index ($A27GG$, solid line) and  $At$ component of the 3D anisotropy ($A27At$, dashed line). $A27GG$ is presented for 1970-2014, $A27At$ - for 1970-2006.}
\end{figure}

\section{Long term changes of the 27-day variation of the GCR anisotropy}
Generally the vector of the GCR solar anisotropy $A$, describing the stream of cosmic rays in the heliosphere, is three dimensional (3D). In this part of paper we investigate the two dimensional (2D) case of the GCR anisotropy vector being the projection of the vector $A$ on the ecliptic plane $(Ar$, $Af)$. Using harmonic analysis method we calculated the amplitudes of the  2D 27-day variations of the GCR anisotropy ($A27A$) \cite{Alania08} for Climax, DeepRiver, Kiel, Moscow and Oulu NMs for each Sun's rotation period during 1965-2014. Amplitudes $A27A$ of the 2D GCR anisotropy for individual NMs are compared with  the  27-day variation of the GCR  anisotropy  vector  (for  $10$  GV  particles)  calculated using Global Spectropgraphic Method (GSM) by the cosmic ray laboratory of IZMIRAN combining all NMs (http://helios.izmiran.troitsk.ru/cosray/main.htm). Fig. ~\ref{fig:1} presents  for the first time the temporal changes of the 27-day variation of the 2D GCR anisotropy $A27A$ smoothed over 13 Sun's rotations for 1965-2014 for Climax, DeepRiver, Kiel, Moscow and Oulu individual NMs and for all combining NMs by GSM. We confirm our previous finding \cite{Alania08} that near the solar minimum the amplitudes of the 27-day variations of the 2D GCR anisotropy determined using GSM basically do not differ from the amplitudes found by harmonic analysis for an individual NM with cut-off rigidity $< 5$ GV. Figure ~\ref{fig:1} shows that this feature is kept for the whole solar cycle and solar magnetic cycle. The time lines of the 27-day variation of the 2D GCR anisotropy  confirm that the average amplitude in the minimum epoch of solar activity is polarity dependent, as it is expected from the drift theory. The amplitude in the negative polarity epochs is less  as we had shown  before \cite{Alania08}. The feeble 11-year variation connected with solar cycle and strong 22-year pattern connected with solar magnetic cycle is visible in the 27-day variation of the 2D anisotropy for 1965-2014.

The next goal is to study the periodic character of the north-south component of the GCR anisotropy. It is performed by two different approaches: (1) using $At$ component obtained by GSM and (2) using $GG$ index being the measure of the north-south anisotropy.
Subsequently using harmonic analysis method we calculated the amplitudes of the 27-day variations of the $GG$ index ($A27GG$) and $At$ component ($A27At$) being the measure of the north-south anisotropy normal to the ecliptic plane. We compare results for the $GG$ index with analysis of the $At$ component obtained by GSM. As far the formulation of the north-south anisotropy is based on the drift model, the $GG$ index being the measure of the north-south asymmetry is highly anticorrelated with $By$ component of the IMF. As an example Fig. ~\ref{fig:2}  presents daily $GG$ index, $By$ component and the sector structure of the IMF in 2008. Figure ~\ref{fig:3} presents time lines of the amplitudes of the 27-day variations of the $GG$ index  and the $At$  component  of the 3D GCR anisotropy for 1970-2014; $A27GG$ is presented for 1970-2014, $A27At$ - for 1970-2006 (data of the $GG$ index starts in 1970 and $At$ component  after 2006 is not available at IZMIRAN website). The general behavior of the 27-day recurrence of the $GG$ index and normal component $At$  of the 3D anisotropy is similar. We show that the 27-day variation of the $GG$ index varies in accordance to solar cycle with a period of 11-years, being in good agreement with the 27-day variation of the $At$  component of the GCR anisotropy. Fig. ~\ref{fig:3}  shows that there is a clear positive correlation between  the changes of  the $A27GG$ and $A27At$, indicating  with a high probability about the  existence of  theirs common source. Thus, data from cosmic ray detectors at low (NM) and high (MT) rigidities confirm no significant dependence of the 27-day variation of the north-south anisotropy upon solar magnetic polarity.

\section{27-day variations of the GCR intensity and 2D anisotropy, GG index, solar wind velocity and IMF components in the Solar Cycle no. 24}

The current Solar Cycle No. 24 (SC 24) is the least active in a Space Age and even in the perspective of a century. The period, from 2007 up to now, because of its very peculiar nature, is broadly investigated in many aspects. One of them is the long-lasting very deep solar minimum 2007-2009 with well established recurrent structures related to the Sun's rotation period ($\sim$27 days). Recently, the peak of the SC 24 has arrived and the second-half of the solar cycle is officially started. Also in the maximum of solar activity in 2014-2015, the 27-day recurrent variations were clearly visible in cosmic ray intensities, $GG$-index, solar wind velocity and heliospheric magnetic field. So the SC 24 offers the possibility to study 27-days recurrent variations over a wide range of conditions: from low to high solar activity.

To reveal  the quasi-periodic effects connected with the rotation of the Sun in the SCs 23 and 24, we employ the wavelet time-frequency spectrum technique developed by Torrence and Compo \cite{TC}. In our calculation we used the Morlet wavelet mother function.
We performed wavelet analysis for the daily solar wind velocity $V$, $By$ component of the IMF, GCR intensity for Oulu NM, $GG$ index, $Ar$ and $Af$ components of the 2D GCR anisotropy during 2007-2008 solar minimum and 2013-2015 solar maximum, considering as a sampling  time interval equaling  two years. In our case a time interval of two years gives a good enough statistics as far our aim is to reveal  recurrences $\leq$ 27-day period, which is $\sim4\%$ from whole sampling period of 26 solar rotations.

\begin{figure}[tbp]
  \begin{center}
\includegraphics[width=0.8\hsize]{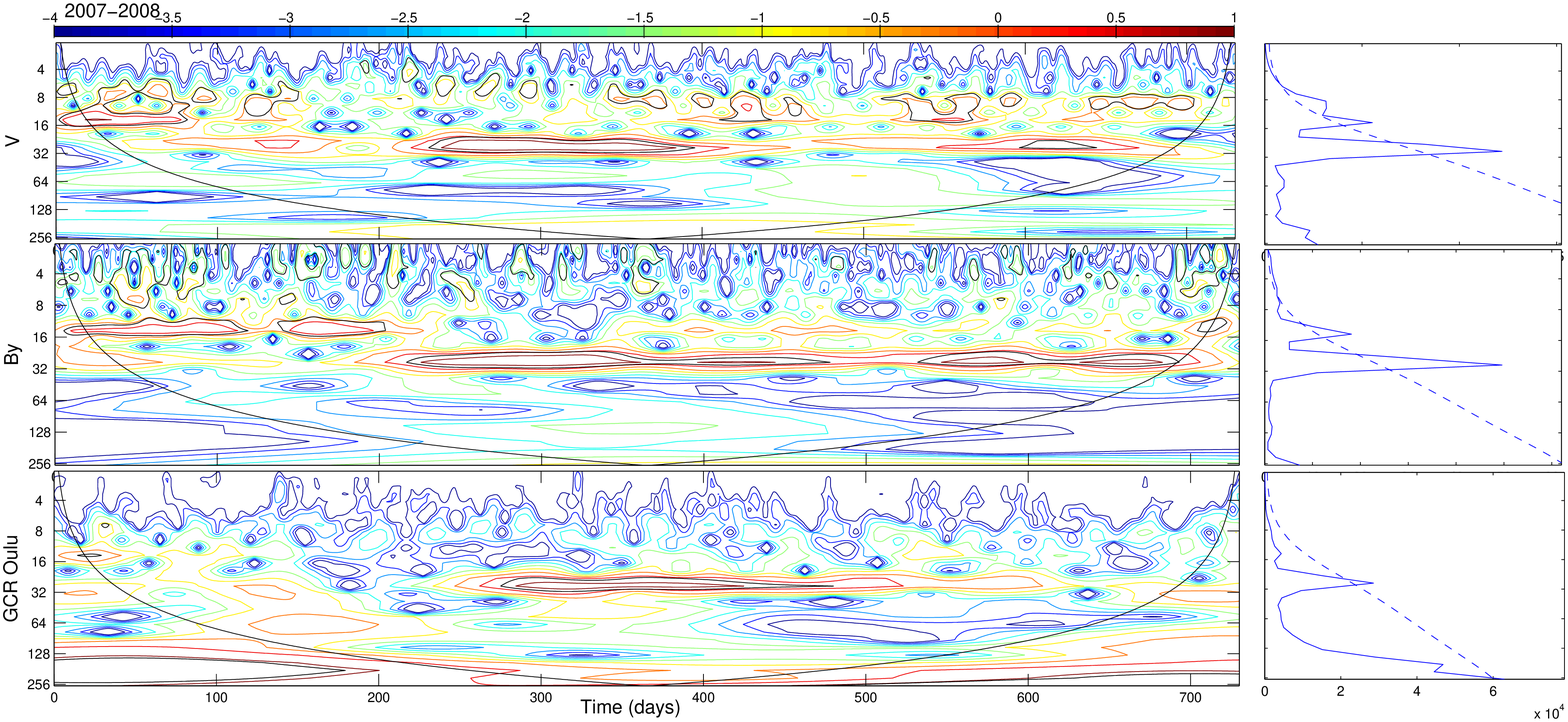}
\end{center}
\caption{\label{fig:4} Wavelet analysis of the daily solar wind velocity $V$ (a), $By$ component of the IMF (b), GCR intensity for Oulu NM (c) for 2007-2008.}
\end{figure}

\begin{figure}[tbp]
  \begin{center}
\includegraphics[width=0.8\hsize]{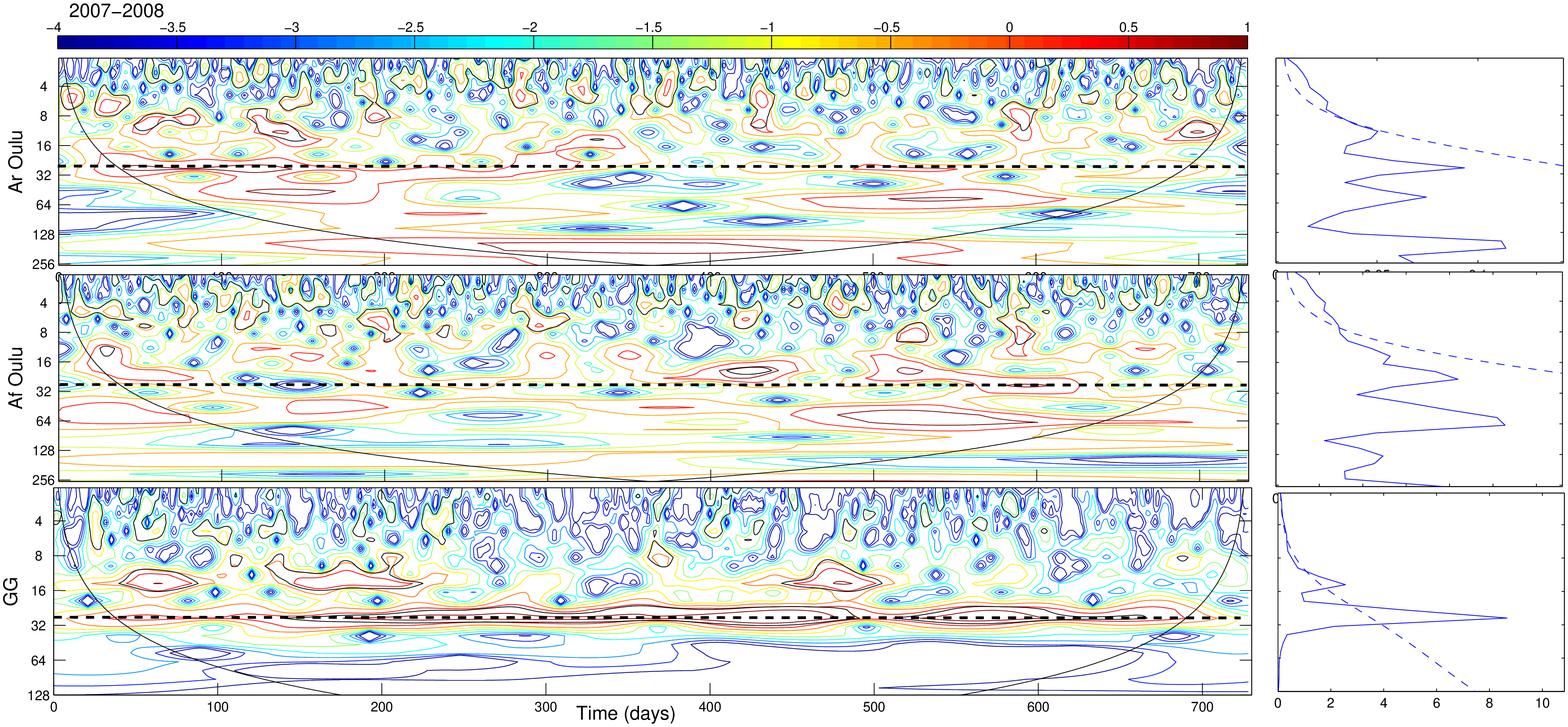}
\end{center}
\caption{\label{fig:5} Wavelet analysis of the $Ar$ (a) and $Af$ (b) components of the 2D GCR anisotropy and daily $GG$ index (c) for 2007-2008. Dashed line designates the period of 27 days.}
\end{figure}

\begin{figure}[tbp]
  \begin{center}
\includegraphics[width=0.8\hsize]{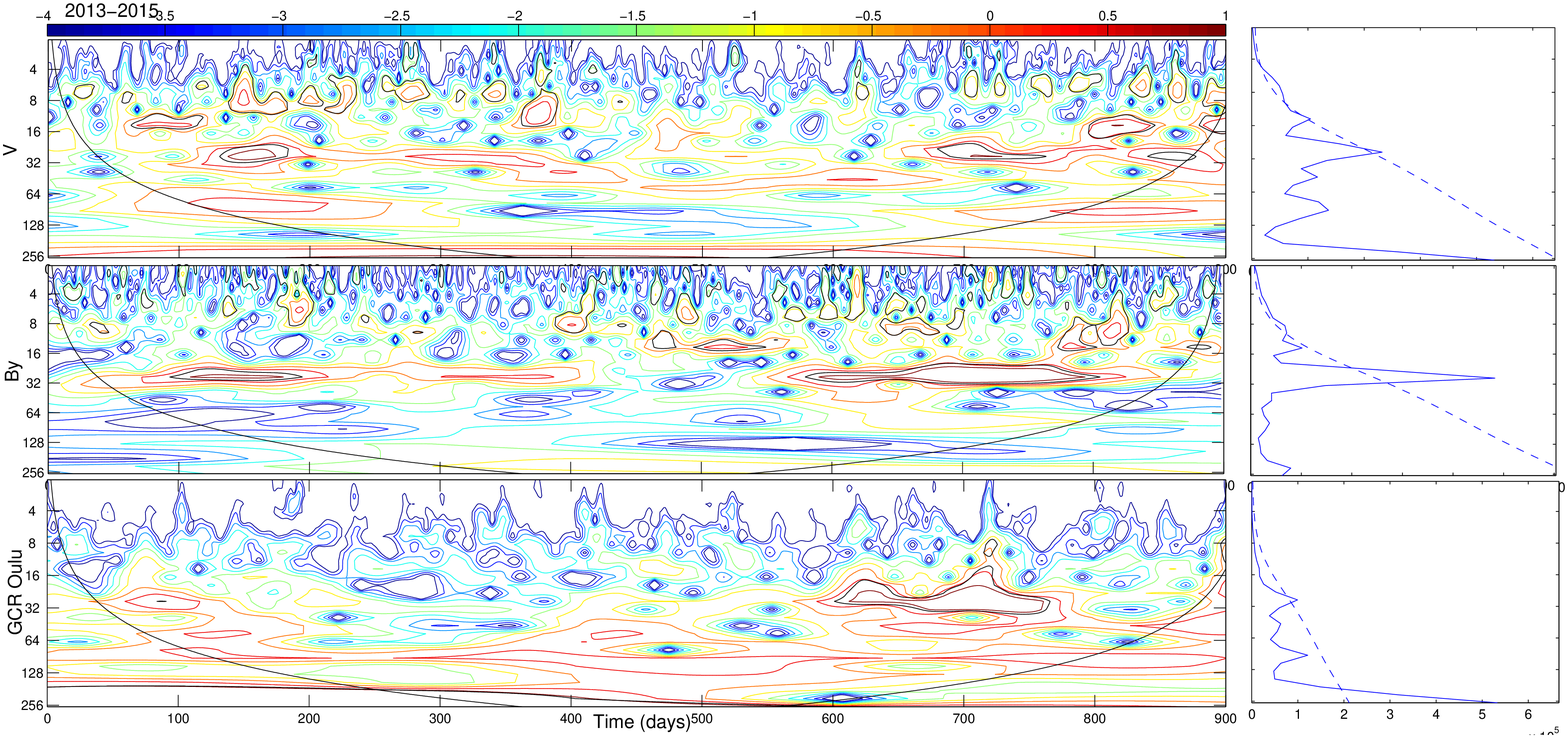}
\end{center}
\caption{\label{fig:6} Wavelet analysis of the daily solar wind velocity $V$ (a), $By$ component of the IMF (b), GCR intensity for Oulu NM (c) for 2013-2015.}
\end{figure}

\begin{figure}[tbp]
  \begin{center}
\includegraphics[width=0.8\hsize]{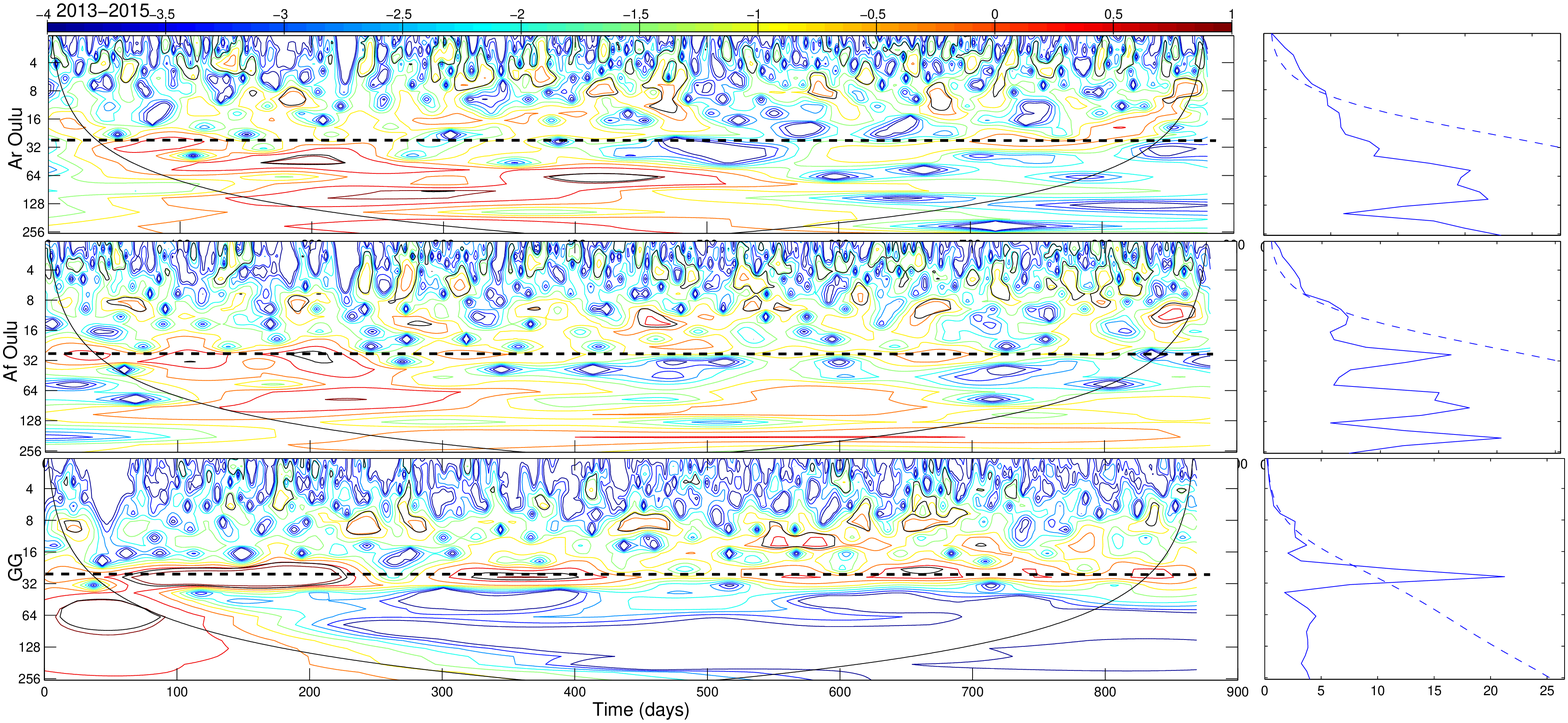}
\end{center}
\caption{\label{fig:7} Wavelet analysis of the $Ar$ (a) and $Af$ (b) components of the 2D GCR anisotropy and daily $GG$ index (c) for 2013-2015. Dashed line designates the period of 27 days.}
\end{figure}

Results of calculation using wavelet analysis method for $V$, $By$ and GCR intensity are presented in Fig. ~\ref{fig:4} (solar minimum)  and  Fig. ~\ref{fig:6} (solar maximum). Results of analysis for $Ar$ and $Af$ components of the 2D GCR anisotropy for Oulu NM and $GG$ index are presented in Fig. ~\ref{fig:5} (solar minimum) and Fig. ~\ref{fig:7} (solar maximum).
Figures ~\ref{fig:5} and ~\ref{fig:7} panel (c) present very clear quasi-periodic changes in $GG$ index related to the Sun's rotation ($\sim 27$ days) for almost whole analyzed time intervals. Similar quasi periodic character is clearly visible in $V$ and $By$ component of the IMF. 27-day variation of GCR intensity is also well established at the end of 2007 and in 2008 during solar minimum (Fig. ~\ref{fig:4}) and at the end of 2014 and in the beginning of 2015 during solar maximum (Fig. ~\ref{fig:6}). Although recurrent variations connected with corotating structures ($\sim 27$ days) are clearly established in almost all solar wind parameters, 2D GCR anisotropy  shows a weak 27-day variation, only in some periods. This is connected with large dispersion of daily $Ar$ and $Af$ components of the 2D GCR anisotropy.

\section{Conclusions}
\begin{enumerate}
  \item The time line of the 27-day variation of the 2D GCR anisotropy  confirm that the average amplitude in the minimum epoch of solar activity is polarity dependent, as it is expected from the drift theory. The amplitude in the negative polarity epochs is less  as we had shown  before. The feeble 11-year variation connected with solar cycle and strong 22-year pattern connected with solar magnetic cycle is visible in the 27-day variation of the 2D anisotropy for 1965-2014.
  \item We show that the 27-day variation of the $GG$ index varies in accordance to solar cycle with a period of 11-years and is almost independent of solar magnetic polarity, being in good correlation with the 27-day variation of the $At$ component of the GCR anisotropy calculated using the GSM method.
  \item We study recurrent variations in 3D GCR anisotropy connected with corotating structures observed in the heliosphere in the Solar Cycles no. 23 and 24. 2D GCR anisotropy generally does not show evident 27-day variation, but in some periods weak recurrent character is visible. This is connected with large dispersion of daily $Ar$ and $Af$ components of the 2D GCR anisotropy. Using wavelet time-frequency method we reveal clear 27-day waves in the $GG$ index, being the measure of the north-south anisotropy, for almost whole analyzed period.

\end{enumerate}

\section*{Acknowledgments}
 We thank providers of data used in this study.


\begin{thebibliography}{99}
\bibitem{Ahluwalia62} H. S. Ahluwalia and A. J. Dessler, \emph{Diurnal variation of cosmic radiation intensity produced by a solar wind}, Planetary and Space Science, {\bf 9}, 195-210, 1962
\bibitem{Krymsky64} G. F. Krymsky, \emph{Diffusion mechanism of daily variation of galactic cosmic ray}, Geomagnetism  and Aeronomy, {\bf 4}, 6, 977-986, 1964
\bibitem{Parker64} E. N. Parker \emph{Theory of streaming of cosmic rays and the diurnal variation}, Planetary and Space Science, {\bf 12} 735-749, 1964
\bibitem{Krymsky66} G. F. Krymski,  A. I. Kuzmin, N. P. Chirkov et al., \emph{Cosmic ray distribution and reception vectors of detectors I}, Geomagnetism  and Aeronomy, {\bf 6}, 991-996, 1966
\bibitem{Krymsky67} G. F. Krymski,  A. I. Kuzmin, N. P. Chirkov et al., \emph{Cosmic ray distribution and reception vectors of detectors II}, Geomagnetism  and Aeronomy, {\bf 7}, 11-16, 1967
\bibitem{Alania08} M. V. Alania, A. Gil and R. Modzelewska, \emph{Study of the 27-day variations of the galactic cosmic ray intensity and anisotropy}, Adv. Space Res.,  \textbf{41}, 280-286, 2008
\bibitem{Swinson69} D. B. Swinson, \emph{'sidereal' cosmic-ray diurnal variations}, Journal of Geophysical Research, {\bf 74}, 5591, 1969
\bibitem{Bieber86}	J. W. Bieber and M. A. Pomerantz, \emph{Solar cycle variation of cosmic-ray north-south anisotropy and radial gradient}, Astrophys. J., {\bf 303}, 843-848, 1986
\bibitem{Bieber93} J. Chen and  J. W. Bieber, \emph{Cosmic-ray anisotropies and gradients in three dimensions}, Astrophys. J., {\bf 405}, 375-389, 1993
\bibitem{MN79} S. Mori and K. Nagashima , \emph{
	Inference of sector polarity of the interplanetary magnetic field from the cosmic ray north-south asymmetry},  Planetary and Space Science, {\bf 27}, 39-46, 1979
\bibitem{Munakata14} K. Munakata, M. Kozai, C. Kato, J. Kota, \emph{Long-term Variation of the Solar Diurnal Anisotropy of Galactic Cosmic Rays Observed with the Nagoya Multi-directional Muon Detector}, Astrophys. J., {\bf 791}, 22,  2014
\bibitem{YK39} S. Yoshida and J. Kondo, J. Geomagn. And Geoelectr., {\bf 6}, 15, 1939
\bibitem{SY92} D. B. Swinson and  S. I. Yasue, \emph{Waves in the cosmic ray north-south anisotropy with periods of 27 days, 1 year, and 11 years}, Journal of Geophysical Research, {\bf 92}, A12, 19 149-19 155, 1992
\bibitem{SF95} D. B. Swinson and Z. Fujii, \emph{27-Day Waves in the Cosmic Ray North-South Anisotropy}, Proceedings of the 24th International Cosmic Ray Conference, \textbf{SH},   576-579, 1995
\bibitem{Y} T. Yeeram, D. Ruffolo, A. Saiz, N. Kamyan and T. Nutaro, \emph{Corotating Solar Wind Structures and Recurrent Trains of Enhanced Diurnal Variation in Galactic Cosmic Rays}, Astrophys. J., {\bf 784}, 136-147, 2014
\bibitem{TC} C. Torrence and G. P. Compo , \emph{A practical guide for wevelet analysis}, Bull. Am. Meteorol. Soc., {\bf 79}, 61-78, 1998

\end{thebibliography}
\end{document}